\documentclass{svjour3}                     % onecolumn (standard format)
\smartqed  % flush right qed marks, e.g. at end of proof
\usepackage{graphicx,amsmath}
\newcommand {\mcu}{\mathcal{U}}

\journalname{Few-Body Systems (FB20)}
\begin{document}

\title{
Universal four-boson system: dimer-atom-atom Efimov effect
and recombination reactions
\thanks{Presented at the 20th International IUPAP Conference on Few-Body Problems in Physics, 20 - 25 August, 2012, Fukuoka, Japan}
}

\titlerunning{Universal four-boson system}        % if too long for running head

\author{A. Deltuva} 
\institute{ A. Deltuva  \at
Centro de F\'{\i}sica Nuclear da Universidade de Lisboa,
P-1649-003 Lisboa, Portugal \\
              \email{deltuva@cii.fc.ul.pt} }

\date{Received: date / Accepted: date}

\maketitle

\begin{abstract}
Recent theoretical developments in the  four-boson  system
with resonant interactions are described. Momentum-space
scattering equations for the four-particle transition operators are used.
The properties of unstable tetramers with approximate dimer-atom-atom
structure are determined. In addition,
 the three- and four-cluster
recombination processes in the four-boson system are studied.
\end{abstract}

\section{Introduction}
\label{intro}

Few-particle systems with resonant interactions
(characterized by a large two-particle scattering length $a$
greatly exceeding the interaction range) 
exhibit universal behaviour. Their  properties 
are independent of the short-range interaction details.
The theoretical study of such systems was initiated by V. Efimov in 1970
who predicted the existence of weakly
bound three-particle states (Efimov trimers)
 with asymptotic discrete scaling symmetry
 \cite{efimov:plb}. The interest in the universal
few-particle systems raised after the Efimov's prediction
got confirmed in the cold-atom physics experiments \cite{kraemer:06a}.
In the present work we summarize some recent theoretical developments 
for the universal system of four identical bosons.

\begin{figure*}[!]
\includegraphics[scale=0.5]{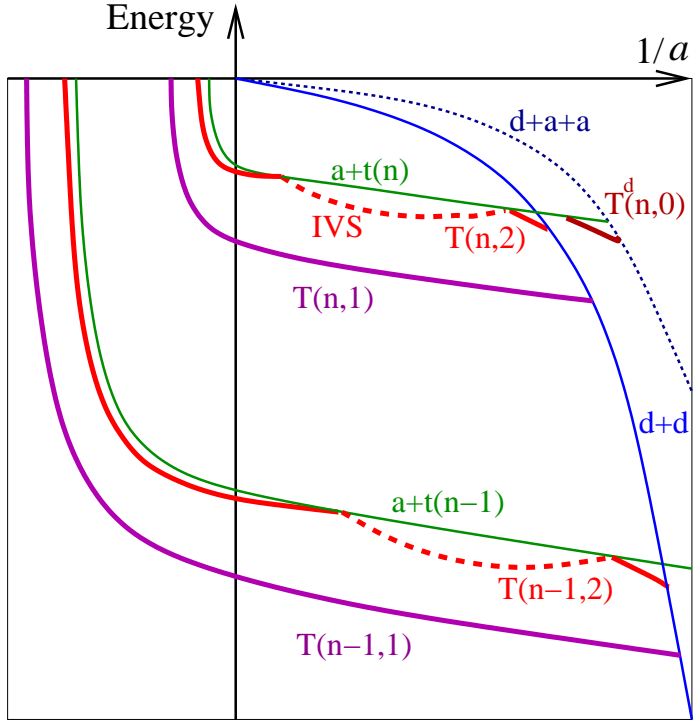} 
\caption{ %(Color online)
 { Schematic representation of the four-boson energy spectrum
as a function of the two-boson scattering length.
The atom-trimer (a+t),  dimer-dimer (d+d),
and dimer-atom-atom (d+a+a) thresholds are shown as thin curves.
The energies for the standard tetramers (T) and for the tetramers
of the dimer-atom-atom structure ($T^d$) are shown as thick curves.
For a better visualization only qualitative but not
quantitative relations between them are preserved and 
only two families of multimers, the $(n-1)$th and $n$th, are shown. 
Positive (negative) $a$ are on the right (left) side from the
vertical axis while the four free atom threshold 
lies on the upper horizontal axis.
The dashed parts of the curves for the shallow tetramers T(n-1,2) and
T(n,2) correspond to the inelastic virtual states (IVS). }}
\label{fig:efi} 
\end{figure*}

The existence of two tetramers
below each Efimov trimer  was predicted in Refs. 
\cite{hammer:07a,stecher:09a}
while their energies and widths were determined very precisely
in Refs. \cite{deltuva:10c,deltuva:11a}. The resulting 
four-boson energy ($E$) spectrum as function of $1/a$ is 
 schematically shown in Fig.~\ref{fig:efi}
 (see refs.~\cite{deltuva:11a,deltuva:12a} for quantitative relations between $a$ and $E$
 that, for a better visualisation, are not preserved here). The above-mentioned
tetramers are labeled as $T(n,m)$ where  $n$ refers to the associated Efimov
trimer and $m=1,2$.
 In the present work we focus on a different type of tetramers,
namely, the ones having an approximate dimer-atom-atom structure
and labeled as $T^d(n,m)$.
Their existence was predicted in Ref.~\cite{braaten:rev} as a
consequence of the three-body Efimov effect in the three-body system made off 
a dimer and two atoms. In a special regime near the 
Efimov trimer intersection with the atom-dimer threshold 
the  atom-dimer  scattering length becomes arbitrarily large
and greatly exceeds the dimer size. Under these circumstances
 one may expect to mimic some properties of the few-boson systems  
using a model that considers the dimer as a pointlike particle.
Then in the effective three-body system consisting of a dimer and two atoms
there are two atom-dimer pairs with infinite two-body scattering length.  
In such a three-body model of the four-boson system
the three-body Efimov effect occurs, however,  with a very large discrete 
scaling factor \cite{braaten:rev}.
The dimer-atom-atom structure of the resulting Efimov-like states
is only an approximation, and we present the first successful attempt 
\cite{deltuva:12b}
to describe them rigorously as four-boson states. Since they lie in the 
continuum, we use scattering equations for the four-particle transition 
operators. 
For  a given Efimov trimer the universal
properties of the lowest associated tetramer $T^d(n,0)$ are determined and
its impact  on the collisions in the four-boson system
is discussed.

Furthermore, using the same technical framework 
we study the recombination processes in the four-boson system,
i.e., the four-boson recombination into atom plus trimer \cite{deltuva:12a}
and dimer-atom-atom recombination into atom plus trimer or dimer plus dimer
\cite{deltuva:12e}.

\section{Four-boson scattering equations \label{sec:4bse}}

We employ 
Alt, Grassberger, and Sandhas (AGS) equations \cite{alt:jinr,grassberger:67} 
for the transition operators that provide an exact description
of the four-particle scattering process.
In the case of four identical  particles there are only 
two distinct two-cluster partitions, one of $3+1$ type and
one of $2+2$ type. We choose those partitions to be (12,3)4 and
(12)(34) and denote them in the following by $\alpha =1$ and $2$,
respectively. The corresponding transition operators $\mcu_{\beta\alpha}$
for the system of four identical bosons obey symmetrized AGS equations 
\begin{subequations} \label{eq:U}
\begin{align}  
\mcu_{11}  = {}&  P_{34} (G_0  t  G_0)^{-1}  
 + P_{34}  U_1 G_0  t G_0  \mcu_{11} + U_2 G_0  t G_0  \mcu_{21} , 
\label{eq:U11} \\  
\mcu_{21}  = {}&  (1 + P_{34}) (G_0  t  G_0)^{-1}  
+ (1 + P_{34}) U_1 G_0  t  G_0  \mcu_{11} , \label{eq:U21} \\
\mcu_{12}  = {}&  (G_0  t  G_0)^{-1}  
 + P_{34}  U_1 G_0  t G_0  \mcu_{12} + U_2 G_0  t G_0  \mcu_{22} , 
\label{eq:U12} \\  
\mcu_{22}  = {}& (1 + P_{34}) U_1 G_0  t  G_0  \mcu_{12} . \label{eq:U22}
\end{align}
\end{subequations}
Here $G_0 = (E+i0-H_0)^{-1}$ is the free Green's function of the
four-particle system with energy $E$ and kinetic energy operator $H_0$,
the two-particle transition matrix $t =v + v G_0 t$ acting within the pair (12)
is derived from the corresponding potential $v$, 
 and the symmetrized operators for the 3+1 and 2+2 subsystems
are obtained from the integral equations
\begin{equation} \label{eq:U3}
U_{\alpha} =  P_\alpha G_0^{-1} + P_\alpha  t G_0  U_{\alpha}.
\end{equation}
 The employed basis states have to be symmetric 
under exchange of two particles in subsystem (12) for the $3+1$ partition
and in (12) and (34) for the $2+2$ partition. 
The correct symmetry of the four-boson system is ensured by the 
operators $P_{34}$, $P_1 =  P_{12}\, P_{23} + P_{13}\, P_{23}$, and
$P_2 =  P_{13}\, P_{24} $ where $P_{ab}$ is the
permutation operator of particles $a$ and $b$.

All observables for two-cluster reactions are determined by the
transition amplitudes
\begin{gather} \label{eq:ampl}
\langle \Phi_{\beta,f}| T |\Phi_{\alpha,i} \rangle  = S_{\beta\alpha}
\langle  \phi_{\beta,f} | \mcu_{\beta\alpha}| \phi_{\alpha,i} \rangle,
\end{gather}
obtained \cite{deltuva:07a} as on-shell matrix elements of 
the  AGS operators \eqref{eq:U}; 
the weight factors $S_{\beta\alpha}$ with values
 $S_{11} = 3$, $S_{22} = 2$, $S_{21} = \sqrt{3}$, and $S_{12} = 2\sqrt{3}$
 arise due to the symmetrization \cite{deltuva:07a}.
The matrix elements \eqref{eq:ampl} are
 calculated between the Faddeev components
\begin{equation} \label{eq:f3}
 | \phi_{\alpha,n} \rangle = G_0 \, t  P_\alpha | \phi_{\alpha,n} \rangle 
\end{equation}
of the corresponding initial/final atom-trimer or dimer-dimer  states
$| \Phi_{\alpha,n} \rangle = (1+P_\alpha)| \phi_{\alpha,n} \rangle$.

The amplitude for the three-cluster breakup of the initial
two-cluster state is given \cite{deltuva:12e} by
\begin{equation} \label{eq:Ud}
 \langle \Phi_{d} |  T_{d \alpha} | \Phi_{\alpha,n} \rangle 
=  S_{d\alpha}  \langle \Phi_{d} | 
[(1+ P_{34}) U_1 G_0 \, t \, G_0 \, \mcu_{1\alpha} + 
U_2 G_0 \,  t \, G_0 \, \mcu_{2\alpha} ]
| \phi_{\alpha,n} \rangle 
\end{equation}
with the symmetrization factors $S_{d1} = \sqrt{3}$ and $S_{d2} = 2$
and the dimer-atom-atom three-cluster channel state $|\Phi_{d}\rangle $.
Due to the time-reversal symmetry the
three-cluster recombination into a two-cluster state is described by the same 
amplitude, i.e.,
$ \langle \Phi_{\alpha,n} |  T_{\alpha d} | \Phi_{d} \rangle =
\langle \Phi_{d} |  T_{d \alpha} | \Phi_{\alpha,n} \rangle $.

The amplitude for the four-cluster breakup can be obtained \cite{deltuva:12a} as
\begin{equation} \label{eq:U0}
\begin{split}  
 \langle \Phi_{0} |  T_{0 \alpha} | \Phi_{\alpha,n} \rangle 
= {}& S_{0\alpha}  \langle \Phi_{0} | (1+P_1)
\{ [1+P_{34}(1+P_1)] 
 t \, G_0    U_1 G_0 \, t \, G_0 \, \mcu_{1\alpha} \\ {}& + 
(1+P_2) t \, G_0    U_2 G_0 \,  t \, G_0 \, \mcu_{2\alpha} \}
| \phi_{\alpha,n} \rangle 
\end{split}
\end{equation}
with $S_{01} = \sqrt{6}$ and $S_{02} = 2\sqrt{2}$ and the
nonsymmetrized four-boson free-channel state $|\Phi_{0} \rangle $.
Again, due to the time-reversal symmetry the amplitude for the
four-boson recombination into a two-cluster state 
is given by 
$  \langle \Phi_{\alpha,n} |  T_{\alpha 0} | \Phi_{0} \rangle
= \langle \Phi_{0} |  T_{0 \alpha} | \Phi_{\alpha,n} \rangle $.

We solve the AGS equations in the momentum-space partial-wave framework;
the technical details can be found in Refs.~\cite{deltuva:12a,deltuva:07a}.

\section{Tetramers of dimer-atom-atom structure \label{sec:tdaa}}

To obtain universal results we
consider reactions involving highly excited Efimov trimers 
where the finite-range effects become negligible; typically,
$n \ge 3$. The $T^d(n,0)$ tetramer emerges at the
$n$th atom-trimer threshold at the two-boson scattering length 
$a = 1.608(1)\, a_n^d$ with $a_n^d$ corresponding to the intersection
of atom-trimer and atom-atom-dimer thresholds.
As shown in  Ref.~\cite{deltuva:12b}, this tetramer stays very close to
the $n$th atom-trimer threshold and decays through the 
atom-atom-dimer threshold at $a \approx 0.999999 \, a_n^d$.
The  higher tetramers $(n,m \ge 1)$ exist only  extremely close to $a=a_n^d$
and cannot be resolved in our four-boson calculations;
estimations based on the three-body model are given in Ref.~\cite{deltuva:12b}.

\begin{figure*}[!]
\includegraphics[scale=0.8]{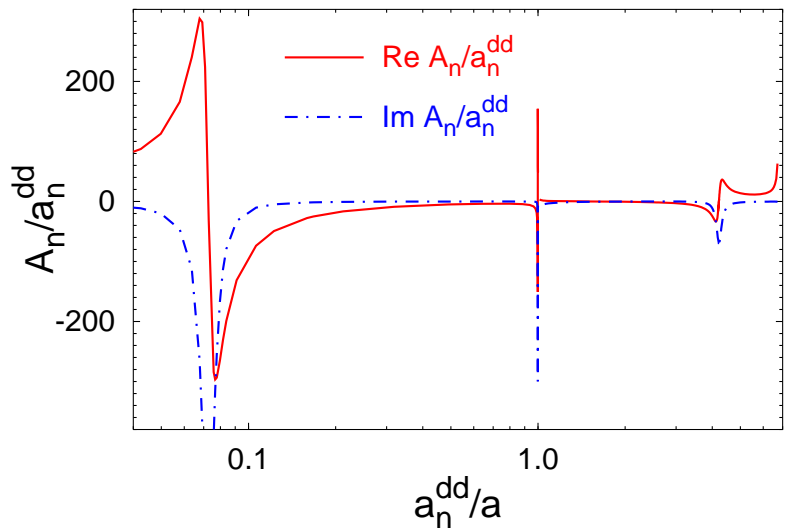}
\caption{Atom-trimer scattering length $A_n$
as function of the two-boson scattering length $a$.}
\label{fig:A}
\end{figure*}

The intersections of the tetramers with the  $n$th atom-trimer threshold
manifest themselves most prominently in the 
collisions of atoms and trimers at vanishing relative kinetic energy
leading to the resonant behaviour of the 
corresponding atom-trimer ($n$th)  scattering length $A_n$.  It is
shown in Fig.~\ref{fig:A} as a function of the 
two-boson scattering length $a$. Here we use $a_{n}^{dd} = 6.789(1)\, a_n^d$
 to build universal dimensionless ratios; this special value of $a$
 corresponds to the intersection of the $n$th atom-trimer and dimer-dimer 
thresholds. The $A_n$ resonance at $a_{n}^{dd}/a  \approx 4.22 $
is due to the $T^d(n,0)$ tetramer.
The increase of $\mathrm{Re}\,A_n$ shown in Fig.~\ref{fig:A}
near  $a_{n}^{dd}/a \approx 6.7$ is not related to the tetramers; it is due to
the scaling of the elastic cross section
with the spatial size of the trimer that grows with decreasing binding
relative to the atom-dimer threshold.

The $A_n$ resonances at $a_{n}^{dd}/a  \approx 0.073 $ and
$a_{n}^{dd}/a  \approx 0.9984 $
correspond to the disappearance and reappearance of the $T(n,2)$ tetramer
as shown in Fig.~\ref{fig:efi}. The explanation for such a spectacular
 behaviour of the $T(n,2)$ tetramer may be the following:
Increasing $1/a$, i.e., increasing the strength of the two-boson interaction,
the binding energies increase for all few-boson states but at different rates.
The increase is faster for the dimer, slower for the trimers, 
and even slower for the tetramers. This is why the $n$th trimer decays through
the atom-dimer threshold at $a=a_n^d$, and the $T(n,2)$ tetramer,
having a structure of an atom weakly attached to the trimer,
decays through the atom-trimer threshold at $a_{n}^{dd}/a \approx 0.073 $,
becoming an inelastic virtual state (IVS). However,
when the the $n$th atom-trimer and dimer-dimer thresholds are 
close to each other, the interference of the atom-trimer and 
dimer-dimer configurations in the $T(n,2)$ tetramer
yield additional binding such that the tetramer becomes an unstable bound
state  until it finally decays through the dimer-dimer threshold 
\cite{deltuva:11b}.

\section{Recombination reactions \label{sec:r}}

\begin{figure*}
\includegraphics[scale=0.58]{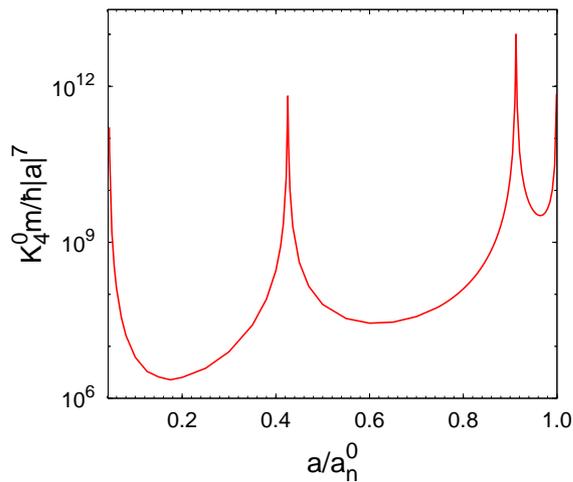}
\caption{\label{fig:k4} 
Ultracold four-boson recombination rate $K_4^0$ as a universal
dimensionless quantity with $m$ being the boson mass.}
%$a/a_n^0$. } %The results are obtained with }
\end{figure*}

The number of the four-atom recombination events 
in a non-degenerated atomic gas 
per volume and time is $K_4 \rho^4/4!$  where $\rho$ is the density of atoms
and $K_4$ is the four-atom recombination rate. Its relation
to the breakup/recombination amplitude is given in Ref.~\cite{deltuva:12a}.
The zero-temperature limit of the four-atom recombination rate
$K_4^0$ is shown in Fig.~\ref{fig:k4} as 
a function of the two-boson scattering length. The special value of
$a = a_n^0 \approx -3.138 \, a_{n}^{dd}$ corresponds to the $n$th Efimov 
trimer being at the three free atom threshold. The two peaks of $K_4^0$ at
$a/a_n^0 \approx 0.4254$ and $a/a_n^0 \approx 0.9125$ 
correspond to the $T(n,1)$ and $T(n,2)$  tetramers being at the
four  free atom threshold, respectively. This is qualitatively
consistent with the results of Ref.~\cite{stecher:09a}.

The results for the dimer-atom-atom recombination reactions leading to 
the atom-trimer or dimer-dimer final states are presented
in Ref.~\cite{deltuva:12e}.

\section{Summary \label{sec:s}}

We studied the dimer-atom-atom three-body Efimov effect using rigorous
four-particle scattering equations for the transition operators
that were solved in the momentum-space framework.
We determined the properties of the deepest resulting tetramers and
demonstrated that in  ultracold atom-trimer collisions 
they lead to resonant enhancement of elastic and inelastic reactions.
Furthermore, we considered three- and four-cluster recombination
reactions in the universal four-boson system and 
presented  results for the ultracold four-boson
recombination. 

%\bibliographystyle{fewbody}
%\bibliography{abbrev,pre80,80-89,90-99,200x,clmb,ad,4N,atomic,numerics}

\end{document}